\documentclass[submission,Phys]{SciPost}
\usepackage[T1]{fontenc}
\usepackage[utf8]{inputenc}
\usepackage{amsmath,amssymb,bm,mathrsfs,graphicx,booktabs,microtype}
\usepackage{amsthm}
\hypersetup{colorlinks=true,allcolors=blue}
\newcommand{\eprint}[2][]{\href{https://arxiv.org/abs/#2}{arXiv:#2}}
\newcommand{\ket}[1]{\lvert#1\rangle}
\newcommand{\bra}[1]{\langle#1\rvert}
\newcommand{\Tr}{\operatorname{Tr}}

\newcommand{\ii}{\mathrm{i}}
\newcommand{\cH}{\mathcal H}
\newcommand{\cD}{\mathcal D}
\newcommand{\cR}{\mathscr R}

\newtheorem{lemma}{Lemma}
\newtheorem{theorem}{Theorem}
\begin{document}
\begin{center}
{\Large\bfseries Wigner--Dyson correlations in selected sectors \\ of an integrable spin chain\par}
\end{center}
\begin{center}
Bal\'azs Pozsgay\textsuperscript{1}
\end{center}
\begin{center}
\textbf{1} MTA-ELTE ``Momentum'' Integrable Quantum Dynamics Research Group,\\
ELTE E\"otv\"os Lor\'and University, Budapest, Hungary
\end{center}

\section*{Abstract}

In quantum many-body physics spectral statistics is often used as a probe to distinguish integrable and chaotic
models. We find a special case of an integrable spin chain with unusual spectral properties. With open boundary
conditions the model has homogeneous exponential degeneracies; these are split for periodic boundary conditions. 
In the latter case, we find that a small number of sectors (selected by a global
symmetry) showcase Wigner-Dyson spectral statistics, which usually signals quantum chaos.
Our model is clean, translationally invariant, and it fits into the standard Yang-Baxter framework. We identify Hilbert space
fragmentation as the key mechanism behind the spectral properties. The fragmentation is non-local, and its extent is
different in the case of open and periodic boundary conditions. We argue that the dynamics of the model is not chaotic 
despite the unusual spectral correlations.

\section{Introduction}
Integrable quantum many-body systems exhibit markedly different physical behavior from generic chaotic systems. For example, their conservation laws generally prevent relaxation to ordinary Gibbs ensembles~\cite{DAlessioKafriPolkovnikovRigol2016,BorgonoviIzrailevSantosZelevinsky2016}, while their large-scale transport can be described by generalized hydrodynamics~\cite{BertiniColluraDeNardisFagotti2016,CastroAlvaredoDoyonYoshimura2016}.

Unlike classical Liouville integrability, however, quantum integrability still lacks a universally accepted general definition, as discussed in detail by Caux and Mossel~\cite{CauxMossel2011}. For quantum spin chains, a standard structural criterion is that the Hamiltonian be obtained as a logarithmic derivative of commuting transfer matrices built from a solution of the Yang--Baxter equation~\cite{Faddeev1996ABA}. Higher logarithmic derivatives generate mutually commuting conserved charges with local densities, giving an infinite hierarchy in the thermodynamic limit.

In parallel, energy-level statistics serve as a widely used diagnostic for distinguishing integrable and chaotic models. After appropriate symmetry resolution, standard integrable chains typically exhibit Poisson statistics, whereas spectra of chaotic models show Wigner--Dyson correlations~\cite{BerryTabor1977,BohigasGiannoniSchmit1984,DAlessioKafriPolkovnikovRigol2016,BorgonoviIzrailevSantosZelevinsky2016}. Caux and Mossel argue that Poisson statistics cannot be used as a universal definition of quantum integrability, while recognizing spectral statistics as one of its physically significant characteristics~\cite{CauxMossel2011}.

Here we construct a homogeneous nearest-neighbor Yang--Baxter-integrable spin chain with exotic dynamical properties. In
our model, the nature of the spectrum depends on the boundary conditions. With open boundaries, the energy levels
coincide with those of the open spin-$\tfrac12$ isotropic Heisenberg (XXX) chain, but each level has an additional
degeneracy that grows exponentially with the chain length. In contrast, with periodic boundaries
we find Wigner--Dyson statistics in selected invariant subspaces (Sec.~\ref{sec:one-statistics}).

Further examination reveals that the model exhibits Hilbert-space fragmentation with both open and
periodic boundaries. For generic choices of the interaction, this fragmentation is nonlocal: the conserved operators
distinguishing the fragments extend across the chain. The extent of the fragmentation is different in the open and the
periodic cases, and it explains the peculiar properties of the spectrum in both cases.

Non-Poisson spectral correlations have been observed in several integrable models,
including bosonic constructions with
interaction order growing with particle number~\cite{BenetLeyvrazSeligman2003}, Hamiltonians built from products of
commuting Gaudin charges~\cite{RelanoDukelskyGomezRetamosa2004}, and correlated-parameter examples in integrable matrix
theory~\cite{ScaramazzaShastryYuzbashyan2016,YuzbashyanShastry2013}. Our setting is distinctive in that we construct a
spin chain with a fixed local dimension and a homogeneous, size-independent nearest-neighbor interaction.

We define the model and its symmetries in Sec.~\ref{sec:model} and establish Yang--Baxter integrability in
Sec.~\ref{sec:YB}. We then compute the open and periodic spectra in Sec.~\ref{sec:spectrum-computation} and examine
their spectral statistics in Sec.~\ref{sec:numerics}, before discussing Hilbert-space fragmentation in
Sec.~\ref{sec:fragmentation}. We present our conclusions in Sec.~\ref{sec:conclusions}. 

\subsection{The use of AI in this work}

We used ChatGPT 6 and 6.1 in all parts of the workflow. The main idea of this work came from the author, after having
finished the related paper \cite{Pozsgay2026SquareRoot}. At the author's request, the AI analytically diagonalized
the integrable Hamiltonians of the form considered here. After studying the general solution, the author recognized
its implications for spectral statistics. The numerical tests were then performed by
the AI. Hilbert space fragmentation was worked out in collaboration between the author and the AI. 

The text was prepared in part by the human author, and in part by the AI under human supervision. The text was reworked
multiple times; the human author reviewed the text paragraph by paragraph, line by line.

\section{Model and global symmetries}\label{sec:model}
We consider a homogeneous quantum spin chain of length $L$ with four states at each site and nearest-neighbor interactions. Each site carries two qubits: one specifies a label, $A$ or $B$, and the other a color, $0$ or $1$. The one-site space is therefore the tensor product
\begin{equation}
 V=\mathbb C^2_{\rm label}\otimes\mathbb C^2_{\rm color}.
\end{equation}
We write its basis states as $\ket{s,a}=\ket{s}_{\rm label}\otimes\ket{a}_{\rm color}$, with $s=A,B$ and $a=0,1$. The same four-dimensional space also admits the direct-sum decomposition
\begin{equation}
 V=V_A\oplus V_B,\qquad
 V_s=\operatorname{span}\{\ket{s,0},\ket{s,1}\},\quad s=A,B.
\end{equation}

We consider both periodic and open boundary conditions. Both Hamiltonians are built from a local gate $R$, which we will define below in Eq.~\eqref{eq:gate}. We denote its action on the ordered pair of sites $(j,j+1)$ by $R_j$. We retain the notation $H$ for the periodic Hamiltonian,
\begin{equation}\label{eq:H}
 H=\sum_{j=1}^{L}R_j,\qquad L+1\equiv1,
\end{equation}
where the closing bond connects sites $L$ and $1$. The open Hamiltonian omits this bond:
\begin{equation}\label{eq:H-open}
 H^{\rm open}=\sum_{j=1}^{L-1}R_j.
\end{equation}

The local gate depends on a unitary matrix $C\in U(4)$ acting on the two color qubits. We leave $C$ unspecified at this stage. It may be chosen arbitrarily, including at random, but the same matrix is used on every bond and at every size.

After grouping the label and color tensor factors of the two sites, we define
\begin{equation}\label{eq:gate}
\begin{split}
 R(C)={}&(\ket{AA}\bra{AA}+\ket{BB}\bra{BB})\otimes I_4\\
 &+\ket{AB}\bra{BA}\otimes C
 +\ket{BA}\bra{AB}\otimes C^\dagger.
\end{split}
\end{equation}
This local gate leaves equal-label pairs unchanged and exchanges unlike labels while acting on their colors. We choose $C$ for the hop $BA\to AB$; the reverse hop applies $C^\dagger$.

Alternatively, in the ordered label sectors $AA,AB,BA,BB$, the local gate has the block-matrix form
\begin{equation}\label{eq:gate-block-matrix}
 R=\begin{pmatrix}
 I_4&0&0&0\\
 0&0&C&0\\
 0&C^\dagger&0&0\\
 0&0&0&I_4
 \end{pmatrix}.
\end{equation}

Every bond preserves the numbers of $A$ and $B$ labels separately. We define the local occupation number operators
\begin{equation}
 n_{s,j}=(\ket{s}\bra{s})_j\otimes I_2,\qquad s=A,B,
\end{equation}
and summing over these occupations gives the total numbers of each label:
\begin{equation}\label{eq:NB}
 N_A=\sum_{j=1}^{L}n_{A,j},\qquad N_B=\sum_{j=1}^{L}n_{B,j}.
\end{equation}
The separate conservation of both label numbers holds for either boundary condition,
\begin{equation}
 [H,N_s]=[H^{\rm open},N_s]=0,\qquad s=A,B.
\end{equation}
Since $N_A+N_B=L$, there is only one independent label-number charge, giving a $U(1)$ symmetry for either boundary condition. Within a fixed-number sector, we also use $N_A,N_B$ for the corresponding integer eigenvalues and denote the sector by $\cH_{L,N_B}$. Its dimension is
\begin{equation}\label{eq:dimension}
 \dim\cH_{L,N_B}=\binom{L}{N_B}2^L.
\end{equation}
For periodic boundary conditions, we denote by $T$ the physical translation operator of the full spin chain. It shifts both the label and the color at every site $j$ to site $j+1$, with $L+1\equiv1$. Since the same interaction acts on every bond of the ring, the periodic Hamiltonian satisfies
\begin{equation}\label{eq:translation-symmetry}
 [T,H]=0.
\end{equation}

The change of basis we use to compute the spectrum in Sec.~\ref{sec:spectrum-computation} follows an approach
used by Kumar~\cite{Kumar2009,Kumar2013} in related spin chains. The specific local gate~\eqref{eq:gate}
considered here was not presented in those works.

Recently Ref. \cite{Pozsgay2026SquareRoot}  studied the dynamics of operator entanglement in a special quantum circuit.
The gate studied in \cite{Pozsgay2026SquareRoot} is a special case of our construction, obtained by choosing the binary
shear $C\ket{a,b}=\ket{a\mathbin{\oplus}b,b}$, where $\oplus$ denotes addition modulo two. In this work we study
Hamiltonian dynamics, whereas \cite{Pozsgay2026SquareRoot} considered only the selected quantum circuit. Nevertheless,
many of our results presented below could be applied also to the model of \cite{Pozsgay2026SquareRoot}. We leave this to
future work.

\section{Yang--Baxter integrability}\label{sec:YB}

We establish integrability of the periodic Hamiltonian~\eqref{eq:H} for every unitary color matrix $C$. Starting from
the algebra of the local gate, we construct a commuting family of transfer matrices whose logarithmic derivatives give
the Hamiltonian and its higher local charges.

Let us examine the properties of the local gate $R$. The homogeneous subspaces $V_A\otimes V_A$ and $V_B\otimes V_B$ are invariant and carry the identity. The two mixed subspaces are exchanged by mutually inverse maps. This two-component organization is a gluing procedure which appeared earlier in the literature~\cite{MajidMarkl1996} and, for suitable set-theoretic gates, is related to twisted unions~\cite{GatevaIvanovaMajid2008,Nichita2013,GomborPozsgay2022}.
When $C$ swaps the two colors, $R(C)$ agrees, up to an additive constant, with the isotropic $(2,2)$ XXC Hamiltonian density of Maassarani~\cite{Maassarani1998XXC} (see also the multiplicity models of Ref.~\cite{Maassarani1999Multiplicity}).
Katsura et al.~\cite{KatsuraMatsuiPalettaPozsgay2025} describe its fragmentation in terms of two conserved color sequences, and Padmanabhan and Korepin~\cite{PadmanabhanKorepin2026} obtain the same density by lifting the spin-$\tfrac12$ XXX interaction.

For \emph{every} unitary $C$, the local gates obey
\begin{equation}\label{eq:localalgebra}
 R_j^\dagger=R_j,\qquad R_j^2=I,\qquad
 R_jR_{j+1}R_j=R_{j+1}R_jR_{j+1}.
\end{equation}
We will prove Hermiticity, involutivity, and the braid relation in Appendix~\ref{app:local-algebra}. No Yang--Baxter constraint on $C$ itself is required; the proof uses unitarity and the presence of only two label types.

We introduce a spectral parameter through the Baxterized gate
\begin{equation}\label{eq:baxter}
 \check\cR(u)=I+uR.
\end{equation}
This matrix satisfies the Yang--Baxter equation, as we will verify in Appendix~\ref{app:local-baxterization}:
\begin{equation}\label{eq:braided}
 \check\cR_{12}(u)\check\cR_{23}(u+v)\check\cR_{12}(v)
 =\check\cR_{23}(v)\check\cR_{12}(u+v)\check\cR_{23}(u).
\end{equation}
To construct a periodic transfer matrix~\cite{Faddeev1996ABA}, let $P$ be the \emph{full four-state} SWAP and define
\begin{equation}\label{eq:transfer}
 \cR(u)=P\check\cR(u),\qquad
 \tau(u)=\Tr_a[\cR_{aL}(u)\cdots\cR_{a1}(u)].
\end{equation}
The ordinary Yang--Baxter equation and the auxiliary-space trace give $[\tau(u),\tau(v)]=0$. Regularity, $\cR(0)=P$, implies $\tau(0)=T$. Differentiating the permutation train gives
\begin{equation}\label{eq:hamderiv}
 H=\tau(0)^{-1}\tau'(0).
\end{equation}

Higher logarithmic derivatives of $T^{-1}\tau(u)$ at $u=0$ generate mutually commuting conserved charges. At derivative orders below the system size, these charges are extensive sums of local densities.

\section{Computation of the spectrum}\label{sec:spectrum-computation}

In this section we determine the spectrum for open and periodic boundary conditions. For the spectral calculation, we regard $A$ as the vacuum label and $B$ as the particle label, and refer to $N_B$ as the particle number. This terminology concerns the label degrees of freedom: both the local vacuum subspace $V_A$ and the local particle subspace $V_B$ carry a color qubit. We denote the full color Hilbert space by $\mathcal F=(\mathbb C^2)^{\otimes L}$, with the tensor factors ordered by physical site. We write $I$ for the identity on the space determined by context.

If $N_A=0$ or $N_B=0$, all labels are identical and every local gate $R_j$ acts as the identity on the full color
space. Thus $H=LI$ and $H^{\rm open}=(L-1)I$ within either sector. Each sector contains $2^L$ color states at the same
energy, giving an exponential degeneracy independently of $C$. 

In the following subsections, we first solve the single-particle sector $N_B=1$ and then treat arbitrary particle number $0<N_B<L$. In each case, we consider both open and periodic boundary conditions.

\subsection{\texorpdfstring{$N_B=1$}{NB=1} with open boundaries}\label{sec:open-one-label}
We determine the eigenvalues and eigenstates of the open Hamiltonian~\eqref{eq:H-open} in the sector $N_B=1$, with $L\geq2$. The trivial case $L=1$ is covered by the uniform-label discussion above.

Let $\ket{x}$ denote the label configuration with the $B$ label at site $x$ and $A$ labels at every other site. Keeping the colors in physical site order, we write a general state as
\begin{equation}\label{eq:open-wave}
 \ket{\Psi}=\sum_{x=1}^{L}\ket{x}\otimes\psi_x,
 \qquad \psi_x\in\mathcal F=(\mathbb C^2)^{\otimes L}.
\end{equation}
Each amplitude $\psi_x$ is a vector in the full color space.

The local gate~\eqref{eq:gate} moves the particle from $x$ to $x+1$ by applying $C_{x,x+1}$ to the two color qubits, and moves it back by applying $C_{x,x+1}^\dagger$. Every equal-label bond contributes the identity. There are $L-3$ such bonds when $1<x<L$, and $L-2$ when the particle is at either endpoint. The Schr\"odinger equation therefore gives
\begin{equation}\label{eq:open-physical-recurrence}
\begin{aligned}
 [E-(L-3)]\psi_x
 &=C_{x-1,x}\psi_{x-1}+C_{x,x+1}^\dagger\psi_{x+1},\\
 [E-(L-2)]\psi_1&=C_{1,2}^\dagger\psi_2,\\
 [E-(L-2)]\psi_L&=C_{L-1,L}\psi_{L-1},
\end{aligned}
\end{equation}
where the first line applies to $2\leq x\leq L-1$.

To solve these equations, we perform a unitary transformation on the color space, and we do this separately for each
particle position. The idea is to ``transform away'' the effect of the color matrix $C$ which accompanies hopping of the labels.

We choose the particle at the 
right endpoint as the reference position and define 
\begin{equation}\label{eq:open-gauge}
 W_L=I,\qquad W_x=C_{x,x+1}^\dagger W_{x+1},\qquad \psi_x=W_x\varphi_x.
\end{equation}
We evaluate the recursion for $x=L-1,\ldots,1$. Each $W_x$ is unitary, and the recursion ensures that the change of basis removes the color matrices from both hopping terms. Substituting $\psi_x=W_x\varphi_x$ into~\eqref{eq:open-physical-recurrence} and multiplying each equation by $W_x^\dagger$ gives
\begin{equation}\label{eq:open-recurrence}
\begin{aligned}
 [E-(L-3)]\varphi_x&=\varphi_{x-1}+\varphi_{x+1},\\
 [E-(L-2)]\varphi_1&=\varphi_2,\\
 [E-(L-2)]\varphi_L&=\varphi_{L-1}.
\end{aligned}
\end{equation}
The first line again holds in the interior. The color components now obey identical scalar equations.

We use the standing-wave Ansatz
\begin{equation*}
 \varphi_x=\cos\!\left[p\left(x-\tfrac12\right)\right]\chi,
 \qquad \chi\in\mathcal F.
\end{equation*}
This Ansatz satisfies the left endpoint equation and, when present, the bulk equation for $E=L-3+2\cos p$. The right endpoint equation then gives $\sin(Lp)\sin(p/2)=0$. Taking $p_n=\pi n/L$ for $n=0,\ldots,L-1$ gives $L$ independent standing waves and the energies
\begin{equation}\label{eq:open-spectrum}
 E_n^{\rm open}=L-3+2\cos\frac{\pi n}{L},
 \qquad n=0,\ldots,L-1.
\end{equation}
Each of these $L$ distinct energies has $2^L$ independent eigenstates, one for each color basis vector $\chi$. Together, the $L2^L$ states span the full $N_B=1$ Hilbert space. The spectrum is independent of $C$, which enters only through the basis changes $W_x$ in the eigenstates.

\subsection{\texorpdfstring{$N_B=1$}{NB=1} with periodic boundaries}\label{sec:real-space-one-label}\label{sec:one-label}

We now close the chain, with $L\geq2$. Applying the open-chain basis change~\eqref{eq:open-gauge} would remove the color gates from the bulk bonds, but would leave a generally nontrivial color holonomy on the closing bond.

Instead of treating this nonlocal holonomy, we exploit translation invariance. Since $[H,T]=0$, we can work at fixed physical momentum. The condition $T^L=I$ then determines the allowed momenta:
\begin{equation}\label{eq:momentum-quantization}
 T\ket{\Psi}=e^{\ii k}\ket{\Psi},
 \qquad k=\frac{2\pi m}{L},\quad m=0,\ldots,L-1.
\end{equation}

As in~\eqref{eq:open-wave}, the one-particle space is $\cH_{L,1}=\mathbb C^L_{\rm position}\otimes\mathcal F$, with $\mathcal F=(\mathbb C^2)^{\otimes L}$ and colors ordered by physical site. We distinguish physical translation from the unitary translation $T_{\rm col}$ of the colors alone:
\begin{equation}\label{eq:color-translation}
 T_{\rm col}\ket{c_1,\ldots,c_L}=\ket{c_L,c_1,\ldots,c_{L-1}}.
\end{equation}
The operator $T_{\rm col}$ acts only on $\mathcal F$ and satisfies $T_{\rm col}^L=I$. Physical translation shifts both the particle and the colors,
\begin{equation*}
 T\bigl(\ket{x}\otimes\chi\bigr)=\ket{x+1}\otimes T_{\rm col}\chi,
\end{equation*}
with site indices understood modulo $L$.

A color vector $\chi$ at particle site $L$ therefore determines the momentum state
\begin{equation}\label{eq:rs-state}
\begin{aligned}
 \ket{\Psi_k(\chi)}
 &=\frac{1}{\sqrt L}\sum_{x=0}^{L-1}
 e^{-\ii kx}T^x\bigl(\ket{L}\otimes\chi\bigr)\\
 &=\frac{1}{\sqrt L}\sum_{x=1}^{L}
 e^{-\ii k(x-L)}\ket{x}\otimes T_{\rm col}^{x-L}\chi.
\end{aligned}
\end{equation}
Expansion of $T$ and periodicity give the second line.

Translation symmetry makes it sufficient to impose the Schr\"odinger equation at site $L$. Substitution of~\eqref{eq:rs-state} gives
\begin{equation}\label{eq:reference-color-equation}
\begin{aligned}
 \bigl[E-(L-2)\bigr]\chi={}&e^{\ii k}C_{L-1,L}T_{\rm col}^\dagger\chi\\
 &+e^{-\ii k}C_{L,1}^\dagger T_{\rm col}\chi.
\end{aligned}
\end{equation}
Let us now define
\begin{equation}\label{eq:explicit-color-transport}
 \cD_L=T_{\rm col}C_{L-1,L}^\dagger.
\end{equation}
This operator first applies the color gate for a leftward hop, then translates the colors to the right. The unitary $\cD_L$ acts on the color space only. It depends on $C$ and $L$ and is independent of the physical momentum $k$.

Homogeneity and color translation give $C_{L,1}^\dagger T_{\rm col}=T_{\rm col}C_{L-1,L}^\dagger$. We can therefore rewrite Eq.~\eqref{eq:reference-color-equation} as $H(k)\chi=E\chi$ on $\mathcal F$, where
\begin{equation}\label{eq:Hk}
 H(k)=(L-2)I+e^{-\ii k}\cD_L+e^{\ii k}\cD_L^\dagger.
\end{equation}
All momentum dependence appears in the explicit phases $e^{\pm\ii k}$.

We choose an orthonormal color eigenbasis of $\cD_L$:
\begin{equation}\label{eq:transport-eigenbasis}
 \cD_L\chi_a=e^{\ii\theta_a}\chi_a,
 \qquad a=1,\ldots,2^L.
\end{equation}
Unitarity makes the phases $\theta_a$ real. Substitution into~\eqref{eq:Hk} gives the energies
\begin{equation}\label{eq:cosine}
 E_{a,k}=L-2+2\cos(\theta_a-k),
\end{equation}
with normalized physical eigenstates $\ket{\Psi_{a,k}}=\ket{\Psi_k(\chi_a)}$ from~\eqref{eq:rs-state}. The same color basis applies at every $k$. Its $2^L$ vectors at each of the $L$ momenta give a complete basis of $L2^L$ states, including when eigenvalues are degenerate.

In Sec.~\ref{sec:periodic-general}, we will extend the calculation to $N_B>1$ and introduce a color holonomy operator associated with the closing bond. For comparison with that calculation, we define its single-particle counterpart as the color transformation accumulated during one full rightward circuit of the particle starting at site $L$:
\begin{equation}\label{eq:root}
 \Omega_L:=\cD_L^{-L}=C_{L-1,L}\cdots C_{1,2}C_{L,1}.
\end{equation}
The equality follows by expanding $\cD_L^{-L}=(C_{L-1,L}T_{\rm col}^\dagger)^L$ using $T_{\rm col}^L=I$. This holonomy acts only on $\mathcal F$ and has eigenvalues $e^{-\ii L\theta_a}$, so different transport phases can give the same winding eigenvalue. If $\Omega_L$ is nondegenerate, its eigenvectors already diagonalize $\cD_L$.

Although there is only one $B$ label, computing the spectrum still requires diagonalizing the nonlocal color unitary $\cD_L$ on $2^L$ states. The Yang--Baxter construction allows any two-site color unitary $C\in U(4)$. We will examine the resulting spectral correlations in Sec.~\ref{sec:one-statistics}.

\subsection{Arbitrary \texorpdfstring{$N_B$}{NB} with open boundaries}\label{sec:open-general}
We now generalize the open-chain transformation~\eqref{eq:open-gauge} to an arbitrary number of particles. We choose the color basis for each label configuration so that every bond acts only on the labels. To make this choice, we must show that different sequences of exchanges leading to the same configuration give the same color basis.

Fix a particle number $0<N_B<L$. A label configuration $w$ is a sequence of $L$ labels $A$ or $B$, ordered by physical site. We consider all such sequences containing $L-N_B$ labels $A$ and $N_B$ labels $B$. The corresponding vectors $\ket w$ span the label Hilbert space $\cH^{\rm label}_{L,N_B}$. Grouping the label and color tensor factors, with the colors kept in physical site order, gives
\begin{equation}\label{eq:open-many-space}
 \cH_{L,N_B}\cong\cH^{\rm label}_{L,N_B}\otimes\mathcal F.
\end{equation}
The label and color spaces have dimensions $\binom{L}{N_B}$ and $2^L$, respectively. A general state has one color amplitude for each label configuration,
\begin{equation}\label{eq:open-many-input}
 \ket\Phi=\sum_w\ket w\otimes\varphi_w,
 \qquad \varphi_w\in\mathcal F.
\end{equation}

We define the change of basis $G$ by its action on product states,
\begin{equation}\label{eq:open-many-gauge}
 G(\ket w\otimes\chi)=\ket w\otimes U_w\chi,
 \qquad \chi\in\mathcal F.
\end{equation}
We construct the color unitaries $U_w$ in~\eqref{eq:open-color-transport} below. By linearity, the physical state is
\begin{equation}\label{eq:open-many-wave}
 \ket\Psi=G\ket\Phi=\sum_w\ket w\otimes\psi_w,
 \qquad \psi_w=U_w\varphi_w.
\end{equation}

Choose the reference configuration $w_0=A^{N_A}B^{N_B}$. Any other configuration can be reached from $w_0$ by adjacent label exchanges. Each exchange is implemented by $R_j$, which swaps the labels at sites $j$ and $j+1$ while applying $I$, $C_{j,j+1}$, or $C_{j,j+1}^\dagger$ to their colors, according to the labels on that bond. For a chosen sequence, define its color map by
\begin{equation}\label{eq:open-color-transport}
 R_{j_r}\cdots R_{j_1}(\ket{w_0}\otimes\chi)
 =\ket w\otimes U_w\chi,
 \qquad \chi\in\mathcal F.
\end{equation}
This map is unitary because every color gate is unitary. The empty sequence gives $U_{w_0}=I$. We will prove in Lemma~\ref{lem:open-color-basis} below that every exchange sequence reaching the same $w$ gives the same $U_w$.

For example, take $L=4$ and $N_B=2$. The sequence $AABB\to ABAB\to ABBA$ moves a $B$ label to the left at each step, first across bond $(2,3)$ and then across bond $(3,4)$. It gives
\begin{equation}\label{eq:open-color-example}
\begin{aligned}
 U_{AABB}&=I,\\
 U_{ABAB}&=C_{2,3}^\dagger,\\
 U_{ABBA}&=C_{3,4}^\dagger C_{2,3}^\dagger.
\end{aligned}
\end{equation}
This example fixes the orientation and order of the gates; the remaining question is whether different exchange paths assign the same basis to a given configuration.

\begin{lemma}[Consistency of the color bases]\label{lem:open-color-basis}
For every label configuration $w$ with $N_A$ labels $A$ and $N_B$ labels $B$, the unitary $U_w$ in~\eqref{eq:open-color-transport} is independent of the chosen sequence of adjacent exchanges from $w_0$ to $w$.
\end{lemma}

\begin{proof}
First consider two exchange sequences representing the same permutation of the sites. The involution and braid relations~\eqref{eq:localalgebra}, together with commutation on disjoint bonds, are the defining relations of the symmetric group $S_L$. Consequently the product of $R_j$ depends only on the permutation, not on its decomposition into adjacent exchanges. Denote this operator by $\rho(\pi)$ for a permutation $\pi$.

Now let $\pi$ and $\pi'$ be two permutations with $\pi w_0=\pi' w_0=w$. Then $h=\pi^{-1}\pi'$ preserves $w_0$. Such permutations form the stabilizer $S_{N_A}\times S_{N_B}$: they permute sites within the reference $A$ block and within the reference $B$ block. These permutations are generated by adjacent exchanges inside the two blocks. Each corresponding $R_j$ acts as the identity on $\ket{w_0}\otimes\mathcal F$, since the neighboring labels are equal.

It follows that $\rho(h)(\ket{w_0}\otimes\chi)=\ket{w_0}\otimes\chi$ for every color vector $\chi$. Substitution of $\pi'=\pi h$ then gives identical results for $\rho(\pi')$ and $\rho(\pi)$ on this entire reference subspace. Hence the two permutations define the same color operator $U_w$. Combined with the independence of the exchange decomposition, this proves the lemma.
\end{proof}

Lemma~\ref{lem:open-color-basis} makes the proposed map~\eqref{eq:open-many-gauge} unambiguously defined. The mutually orthogonal label configurations and the unitarity of every $U_w$ then make $G$ unitary on the full fixed-$N_B$ space. It preserves the label configuration while changing its associated color basis. Because this basis depends on the full configuration, $G$ is in general nonlocal.

For $N_B=1$, the reference configuration $w_0=A^{L-1}B$ has the particle at site $L$, as in~\eqref{eq:open-gauge}. Moving it left to site $x$ applies the same sequence of color gates as the recursion defining $W_x$. Thus the one-particle restriction is exactly
\begin{equation}\label{eq:open-gauge-comparison}
 U_{A^{x-1}BA^{L-x}}=W_x.
\end{equation}

The purpose of introducing $G$ for arbitrary $N_B$ is to transform each physical exchange into a pure label exchange. Let $P^{\rm label}_{j,j+1}$ swap the neighboring $A/B$ labels, acting as the identity when they are equal. The following result identifies the resulting dynamics.

\begin{theorem}[Open XXX chain]\label{thm:open-XXX}
For any $C\in U(4)$ and fixed $0<N_B<L$, the transformation~\eqref{eq:open-many-gauge} gives
\begin{equation}\label{eq:open-XXX}
\begin{split}
 G^\dagger H^{\rm open}G
 &=H^{\rm XXX,open}_{L,N_B}\otimes I,\\
 H^{\rm XXX,open}_{L,N_B}
 &=\sum_{j=1}^{L-1}P^{\rm label}_{j,j+1}\big|_{\cH^{\rm label}_{L,N_B}}.
\end{split}
\end{equation}
Here $H^{\rm XXX,open}_{L,N_B}$ is the open spin-$\tfrac12$ XXX Hamiltonian restricted to $N_B$ particles, in the permutation normalization.
\end{theorem}

\begin{proof}
Consider one bond $1\leq j<L$, and let $s_jw$ denote the configuration obtained by swapping its two labels. Appending this exchange to a path from $w_0$ to $w$ produces a path to $s_jw$. Lemma~\ref{lem:open-color-basis} therefore gives
\begin{equation}\label{eq:open-bond-transport}
 R_j(\ket w\otimes U_w\chi)
 =\ket{s_jw}\otimes U_{s_jw}\chi.
\end{equation}
For unlike labels, this identity says that the physical color gate converts the basis $U_w$ into the basis $U_{s_jw}$. For instance, the second hop in~\eqref{eq:open-color-example} satisfies $C_{3,4}^\dagger U_{ABAB}=U_{ABBA}$. For equal labels, $s_jw=w$ and $R_j$ acts as the identity, so the same identity holds.

Applying $G^\dagger$ to~\eqref{eq:open-bond-transport} removes $U_{s_jw}$. Since $w$ and $\chi$ span the fixed-$N_B$ space, we obtain the operator identity
\begin{equation}\label{eq:bulk}
 G^\dagger R_jG=P^{\rm label}_{j,j+1}\otimes I.
\end{equation}
Summing this identity over all $L-1$ open-chain bonds proves~\eqref{eq:open-XXX} and completes the proof.
\end{proof}

We now discuss the consequences of this unitary transformation. Every eigenstate of the open XXX label chain can be paired with any of the $2^L$ independent color states. Applying $G$ gives a complete physical eigenbasis, so the energies are independent of $C$. An XXX eigenvalue with multiplicity $d$ has multiplicity exactly $d\,2^L$ in the full fixed-$N_B$ space. At $N_B=1$, these energies reproduce~\eqref{eq:open-spectrum}.

The transformed color degrees of freedom carry no dynamics: time evolution acts only on the label factor in~\eqref{eq:open-XXX}. Physical color observables can nevertheless evolve, because their transformed operators generally depend on the labels through $G$. The gate $C$ thus affects the physical eigenstates and observables even though it does not affect the open-chain energies.

In Sec.~\ref{sec:periodic-general}, we will add the closing bond and show that it leaves a generally nontrivial color operator at the boundary.

\subsection{Arbitrary \texorpdfstring{$N_B$}{NB} with periodic boundaries}\label{sec:periodic-general}
We now close the chain and ask how the colors affect its spectrum. Keep $0<N_B<L$, with $N_A=L-N_B$. We group the bonds $j<L$ into $H^{\rm open}$ and apply the same unitary $G$ as for the open chain. Using~\eqref{eq:open-XXX}, we obtain the similarity transformation of the periodic Hamiltonian:
\begin{equation}\label{eq:periodic-open-seam}
\begin{aligned}
 G^\dagger HG
 &=G^\dagger(H^{\rm open}+R_L)G\\
 &=H^{\rm XXX,open}_{L,N_B}\otimes I
   +G^\dagger R_LG.
\end{aligned}
\end{equation}
The first term is the known open XXX label Hamiltonian, so only the transformed closing bond $G^\dagger R_LG$ remains to be computed. We will first compute the accompanying color map for the reference configuration and then show in Lemma~\ref{lem:periodic-seam} below that it is independent of the interior label configuration. Its eigenvalues will then determine the scalar boundary phases of the XXX chains in Theorem~\ref{thm:periodic-XXX} below. We will subsequently determine how physical translation acts in these blocks and introduce their Bethe equations.

The closing bond exchanges opposite labels at sites $L$ and $1$, leaving the interior labels fixed. To describe the nontrivial exchange, write $v=s_2\cdots s_{L-1}$ for an interior label configuration with $N_A-1$ labels $A$ and $N_B-1$ labels $B$. Thus $AvB$ has label $A$ at site $1$ and label $B$ at site $L$, while $BvA$ has the same interior and the opposite endpoint labels.

Starting from $\ket{AvB}\otimes\chi$, the transformation $G$ gives the color amplitude $U_{AvB}\chi$. The ordered closing pair $(L,1)$ has labels $BA$, so $R_L$ exchanges them and applies $C_{L,1}$ to the colors. Finally, $G^\dagger$ applies $U_{BvA}^\dagger$, the inverse basis change for the final label configuration. The transformed bond therefore acts as
\begin{equation}\label{eq:periodic-seam-color-map}
\begin{aligned}
 G^\dagger R_LG(\ket{AvB}\otimes\chi)
 &=\ket{BvA}\otimes \Omega_v\chi,\\
 \Omega_v&=U_{BvA}^\dagger C_{L,1}U_{AvB}.
\end{aligned}
\end{equation}
We first evaluate $\Omega_v$ for the reference configuration $w_0=A^{N_A}B^{N_B}$, whose interior is $v_0=A^{N_A-1}B^{N_B-1}$. We define the color holonomy $\Omega=\Omega_{L,N_B}$ by $\Omega:=\Omega_{v_0}$. For this configuration, $U_{w_0}=I$. First cross the closing bond, applying $C_{L,1}$ and producing the label configuration $BA^{N_A-1}B^{N_B-1}A$. Return the labels to the reference by moving the final $A$ left through the $B$ block, across bonds $L-1$ down to $N_A+1$, and then moving the initial $B$ right across bonds $1$ through $N_A$. By the path independence proved in Lemma~\ref{lem:open-color-basis}, these return exchanges implement the inverse color basis change for the configuration reached after crossing the closing bond. The closed sequence therefore computes $\Omega_{v_0}=\Omega$, giving
\begin{equation}\label{eq:Omega}
\begin{split}
 \Omega={}&(C_{N_A,N_A+1}\cdots C_{1,2})\\
 &\times(C_{N_A+1,N_A+2}\cdots C_{L-1,L})C_{L,1}.
\end{split}
\end{equation}
An empty product is the identity. The closed sequence contains $1+(N_B-1)+N_A=L$ color gates. For example, at $L=4$, $N_B=2$, the sequence $AABB\to BABA\to BAAB\to ABAB\to AABB$ gives $\Omega_{4,2}=C_{2,3}C_{1,2}C_{3,4}C_{4,1}$.

\begin{samepage}
\begin{lemma}[Closing bond]\label{lem:periodic-seam}
For every allowed interior label configuration $v$, the color operator $\Omega_v$ in~\eqref{eq:periodic-seam-color-map} equals the unitary $\Omega$ given by~\eqref{eq:Omega}.
\end{lemma}
\end{samepage}

\begin{proof}
Each $\Omega_v$ is unitary because it is a product of unitary color maps. It remains to prove its independence of $v$.

For $2\leq j\leq L-2$, the physical bonds $R_j$ and $R_L$ act on disjoint sites and commute. Conjugation by $G$ and~\eqref{eq:bulk} show that $G^\dagger R_LG$ commutes with every adjacent exchange of the interior labels. In the transformed basis, such an interior exchange changes $v$ but leaves $\chi$ unchanged. Commuting it past the closing bond therefore gives $\Omega_v=\Omega_{v'}$ whenever $v'$ is obtained from $v$ by one adjacent interior exchange.

These exchanges connect all configurations $v$ with the prescribed counts, so $\Omega_v=\Omega_{v_0}=\Omega$. If there is only one possible interior configuration, it is $v_0$ and the equality holds by definition. This proves the lemma.
\end{proof}

Hermiticity determines the reverse exchange:
\begin{equation}\label{eq:periodic-seam-reverse}
 G^\dagger R_LG\bigl(\ket{BvA}\otimes\chi\bigr)
 =\ket{AvB}\otimes\Omega^\dagger\chi.
\end{equation}

So far, we have computed the transformed bond only for opposite endpoint labels. The remaining configurations have the form $AvA$ or $BvB$, with the interior labels chosen to give the fixed total particle number $N_B$. On these configurations, $R_L$ acts as the identity on both labels and colors. Since $G$ preserves the label configuration and $U_w^\dagger U_w=I$, the transformed bond $G^\dagger R_LG$ also acts as the identity.

We now combine these results to obtain the full periodic Hamiltonian in the basis defined by $G$. We use the label operators $\sigma^- =\ket B\bra A$, $\sigma^+=(\sigma^-)^\dagger$, and $\sigma^z=\ket A\bra A-\ket B\bra B$. Then
\begin{equation}\label{eq:sortedH}
 G^\dagger HG=\sum_{j<L}P^{\rm label}_{j,j+1}\otimes I
 +\frac{I+\sigma_1^z\sigma_L^z}{2}\otimes I
 +\sigma_1^-\sigma_L^+\otimes\Omega
 +\sigma_1^+\sigma_L^-\otimes\Omega^\dagger.
\end{equation}
The first term gives the XXX exchanges along the open chain. The second term is the identity on equal endpoint labels
and vanishes on unlike endpoint labels. The last two terms exchange unlike endpoint labels, applying $\Omega$
for $AvB\to BvA$ and $\Omega^\dagger$ for the reverse exchange. Thus the bulk remains the XXX label chain, while
the color dependence is confined to the closing bond. The operator we obtained can be interpreted as an XXX chain with
an operator-valued twist.

Because the same color operator accompanies every boundary exchange in a given direction, choosing its eigenbasis
converts the operator boundary condition into scalar boundary phases (scalar twists). The following theorem states the resulting
decomposition, including all multiplicities. 

\begin{theorem}[Periodic XXX blocks]\label{thm:periodic-XXX}
Let $\Omega\chi_a=\xi_a\chi_a$, with an orthonormal color basis $\{\chi_a\}_{a=1}^{2^L}$. The periodic Hamiltonian has the unitary decomposition
\begin{equation}\label{eq:directsum}
 H|_{\cH_{L,N_B}}\simeq\bigoplus_{a=1}^{2^L}H^{\rm XXX}_{L,N_B}(\xi_a),
\end{equation}
where $H^{\rm XXX}_{L,N_B}(\xi_a)$ is obtained from~\eqref{eq:sortedH} by replacing $\Omega$ with $\xi_a$ and $\Omega^\dagger$ with $\xi_a^{-1}$.
\end{theorem}

\begin{proof}
Fix a color eigenvector $\chi_a$ and let $\ket{\phi}$ be any label state. On $\ket{\phi}\otimes\chi_a$, substituting the color eigenvalue into~\eqref{eq:sortedH} gives the label Hamiltonian $H^{\rm XXX}_{L,N_B}(\xi_a)$ and leaves the color channel invariant. These mutually orthogonal channels span all $\binom{L}{N_B}2^L$ states. Applying the unitary $G$ gives the complete physical decomposition~\eqref{eq:directsum}, including all multiplicities.
\end{proof}

Similar mechanisms occur in other models. For example, the Jordan--Wigner transformation of a periodic XY chain introduces the global fermion-parity operator into the term connecting the two ends of the chain~\cite{LiebSchultzMattis1961}. In each parity sector, this operator becomes a scalar boundary sign. In our model, the boundary operator acts on the full color space, and its eigenvalues $\xi_a$ supply the boundary phases of the XXX blocks.

To assign the XXX eigenstates to physical momentum sectors, we now compute the transformed translation $\widetilde T:=G^\dagger TG$, keeping track of the color basis changes that accompany a translation of the labels. The allowed momenta remain $k=2\pi m/L$, since $T^L=I$.

Let $T_{\rm lab}$ cyclically shift the labels one site to the right, and recall the color translation $T_{\rm col}$ defined in Eq.~\eqref{eq:color-translation}. On the grouped label--color space~\eqref{eq:open-many-space}, physical translation factorizes as
\begin{equation}\label{eq:translation-factorization}
 T=T_{\rm lab}\otimes T_{\rm col}.
\end{equation}
For each configuration $w$, set $w'=T_{\rm lab}w$. To compute $\widetilde T$, we apply the three operators in turn. First, $G$ maps the color vector to $U_w\chi$, as in~\eqref{eq:open-many-gauge}. Physical translation then shifts the label configuration to $w'$ and the color vector to $T_{\rm col} U_w\chi$. Finally, $G^\dagger$ applies the inverse color basis change associated with the translated configuration. Thus
\begin{equation}\label{eq:configuration-translation-color}
\begin{aligned}
 \widetilde T(\ket w\otimes\chi)
 &=\ket{w'}\otimes K_w\chi,\\
 K_w&=U_{w'}^\dagger T_{\rm col} U_w.
\end{aligned}
\end{equation}
We first evaluate this color map for $w=A^{N_A-1}B^{N_B}A$, whose translate is the reference configuration $w_0=A^{N_A}B^{N_B}$. Since $U_{w_0}=I$, we have $K_w=T_{\rm col} U_w$. To obtain $w$ from $w_0$, we move the last $A$ of the reference $A$ block to the right through all $N_B$ labels $B$. These exchanges give $U_w=C_{L-1,L}^\dagger\cdots C_{N_A,N_A+1}^\dagger$, and hence the color transport operator $\cD=\cD_{L,N_B}$ takes the form
\begin{equation}\label{eq:general-color-transport}
 \cD:=T_{\rm col} U_w
 =T_{\rm col} C_{L-1,L}^\dagger\cdots C_{N_A,N_A+1}^\dagger.
\end{equation}
Thus $\cD$ is the color transformation accompanying one physical translation of this configuration.

In Appendix~\ref{app:periodic-translation}, we will show that $K_w$ depends only on whether $w$ ends in $A$ or $B$. The reference calculation above therefore gives its value for every configuration ending in $A$. For a configuration ending in $B$, the translation carries an additional color holonomy across the closing bond. The two cases are
\begin{equation}\label{eq:translation-color-cases}
 K_w=
 \begin{cases}
  \cD, & \text{$w$ ends in $A$},\\
  \Omega\cD, & \text{$w$ ends in $B$}.
 \end{cases}
\end{equation}
The same appendix will also prove
\begin{equation}\label{eq:general-transport-relations}
 [\cD,\Omega]=0,
 \qquad \cD^L\Omega^{N_B}=I.
\end{equation}
The second identity expresses $T^L=I$: over $L$ translations, each step contributes $\cD$ and each of the $N_B$ labels $B$ crosses the closing bond once, contributing $\Omega$.

The commuting color unitaries admit a common orthonormal eigenbasis. We refine the basis of Theorem~\ref{thm:periodic-XXX}, including within degenerate eigenspaces of $\Omega$, so that
\begin{equation}\label{eq:joint-color-eigenbasis}
 \Omega\chi_a=\xi_a\chi_a,
 \qquad \cD\chi_a=\eta_a\chi_a.
\end{equation}
In color channel $a$, every translation therefore contributes a phase $\eta_a$, with an additional phase $\xi_a$ when a $B$ label crosses from site $L$ to site $1$. To express this boundary phase, let $n^{\rm label}_{B,L}=(\ket B\bra B)_L$ and define the twisted label translation
\begin{equation}\label{eq:twisted-label-translation}
 T_{{\rm lab},\xi}=T_{\rm lab}\,[I+(\xi-1)n^{\rm label}_{B,L}].
\end{equation}
Substitution of the color eigenvalues into~\eqref{eq:translation-color-cases} gives the transformed physical translation in this channel:
\begin{equation}\label{eq:translation-color-channel}
 \widetilde T(\ket{\phi}\otimes\chi_a)
 =\eta_a(T_{{\rm lab},\xi_a}\ket{\phi})\otimes\chi_a.
\end{equation}
Since $H$ and $T$ commute, we can choose physical energy eigenstates with definite momentum. For an eigenstate $\ket\Psi=G(\ket{\phi}\otimes\chi_a)$ satisfying $T\ket\Psi=e^{\ii k}\ket\Psi$, the left-hand side of~\eqref{eq:translation-color-channel} equals $e^{\ii k}\ket{\phi}\otimes\chi_a$. Comparing the two sides gives
\begin{equation}\label{eq:physical-momentum-twist}
 T_{{\rm lab},\xi_a}\ket{\phi}=e^{\ii k}\eta_a^{-1}\ket{\phi}.
\end{equation}
This condition selects the XXX eigenstates in color channel $a$ that have physical momentum $k$. Each $B$ crosses the closing bond once under $L$ label translations, so $\bigl(T_{{\rm lab},\xi_a}\bigr)^L=\xi_a^{N_B} I$. Applying~\eqref{eq:physical-momentum-twist} $L$ times and using~\eqref{eq:general-transport-relations} therefore gives
\begin{equation}\label{eq:physical-momentum-quantization-check}
 e^{\ii kL}=\eta_a^L\xi_a^{N_B}=1.
\end{equation}
This agrees with the physical momentum quantization~\eqref{eq:momentum-quantization}.

For $1<N_B<L-1$, fixing the physical momentum generally leaves nontrivial label dynamics associated with the relative particle positions. We can compute the label spectra using the Bethe Ansatz for the scalar-twist XXX blocks. Its standard XXX construction~\cite{Faddeev1996ABA} describes the $N_B$ label particles by rapidities $\lambda_1,\ldots,\lambda_{N_B}$. Define their plane-wave factors by
\begin{equation}\label{eq:Bethe-wave-factor}
 z_j=\frac{\lambda_j+\ii/2}{\lambda_j-\ii/2}.
\end{equation}
Taking one particle around the ring accumulates its propagation factor $z_j^L$, the two-particle scattering factors of the XXX chain, and the boundary phase. For the closing bond in~\eqref{eq:sortedH}, a single-particle wave $z^x$ obeys $z^L=\xi_a^{-1}$. The resulting equations for regular Bethe states are
\begin{equation}\label{eq:BAE}
 \left(\frac{\lambda_j+\ii/2}{\lambda_j-\ii/2}\right)^L
 =\xi_a^{-1}\prod_{b\ne j}
 \frac{\lambda_j-\lambda_b+\ii}{\lambda_j-\lambda_b-\ii}.
\end{equation}
The left side is the propagation around the ring; the product on the right contains the scattering factors, and $\xi_a^{-1}$ is the boundary contribution. The energies in the permutation normalization are
\begin{equation}\label{eq:energyBAE}
 E=L-\sum_{j=1}^{N_B}\frac{1}{\lambda_j^2+1/4}.
\end{equation}
Common translation of the coordinates multiplies every coordinate Bethe wave by $\prod_jz_j$. The operator $T_{{\rm lab},\xi_a}$ therefore has eigenvalue $(\prod_jz_j)^{-1}$, including configurations crossing the boundary. Substitution into~\eqref{eq:physical-momentum-twist} gives
\begin{equation}\label{eq:Bethe-physical-momentum}
 e^{\ii k}=\eta_a\prod_{j=1}^{N_B}
 \frac{\lambda_j-\ii/2}{\lambda_j+\ii/2}.
\end{equation}
This condition selects the physical momentum within the XXX decomposition. Multiplication of the Bethe equations gives $(\prod_jz_j)^L=\xi_a^{-N_B}$, consistent with the translation identities above.

Finally, set $N_B=1$ to compare with Sec.~\ref{sec:real-space-one-label}. Equation~\eqref{eq:general-color-transport} becomes $\cD_{L,1}=\cD_L$, and~\eqref{eq:general-transport-relations} gives $\Omega_{L,1}=\Omega_L=\cD_L^{-L}$, as in~\eqref{eq:root}. On a joint color eigenvector, the boundary phase and the momentum condition are
\begin{equation}\label{eq:periodic-one-particle-check}
 \eta_a=e^{\ii\theta_a},\qquad
 \xi_a=e^{-\ii L\theta_a},\qquad
 z=e^{\ii(\theta_a-k)}.
\end{equation}
For one rapidity, $1/(\lambda^2+1/4)=2-z-z^{-1}$. Substitution into~\eqref{eq:energyBAE} therefore gives $E=L-2+2\cos(\theta_a-k)$, exactly~\eqref{eq:cosine}. To compare the eigenstates, the recursion~\eqref{eq:open-gauge} and $\cD_L\chi_a=\eta_a\chi_a$ give $W_x\chi_a=\eta_a^{L-x}T_{\rm col}^{x-L}\chi_a$. Consequently the physical amplitudes satisfy
\begin{equation}\label{eq:periodic-one-particle-state-check}
 W_xz^{x-L}\chi_a
 =e^{-\ii k(x-L)}T_{\rm col}^{x-L}\chi_a.
\end{equation}
Summing these amplitudes with normalization $L^{-1/2}$ reproduces precisely the momentum state~\eqref{eq:rs-state}. Thus both constructions give the same energies and physical eigenstates in a common color eigenbasis.

\section{Numerical spectral statistics}\label{sec:numerics}
In this section we investigate the level-spacing statistics of the periodic model in sectors of fixed particle number and momentum, comparing them with Poisson and unitary-ensemble references for one fixed color matrix $C$. Instead of diagonalizing the full spin-chain Hamiltonian, we use the spectral decomposition established in Sec.~\ref{sec:periodic-general}: the energies are obtained from spin-$\tfrac12$ XXX chains with scalar boundary twists, with the twist parameters given by the eigenvalues of the unitary operator $\Omega_{L,N_B}$ acting on the color space. The common color eigenbasis~\eqref{eq:joint-color-eigenbasis} also supplies the transport phases needed to select physical momentum through~\eqref{eq:physical-momentum-twist}. For small systems, we verified the reconstructed spectra by directly diagonalizing the physical Hamiltonian in fixed momentum sectors.

We generate the color unitary $C$ by applying a Cayley transform to a Hermitian matrix $M$. For all numerical calculations, we use the following fixed choice in the basis $\ket{00},\allowbreak\ket{01},\allowbreak\ket{10},\allowbreak\ket{11}$ of the ordered color pair:
\begin{equation}\label{eq:numerical-gate}
 C=(7I_4-\ii M)(7I_4+\ii M)^{-1},
\end{equation}
where
\begin{equation}\label{eq:numerical-hermitian-matrix}
 M=\begin{pmatrix}
 2&3+2\ii&-1+4\ii&2-3\ii\\
 3-2\ii&-3&1-2\ii&-4+\ii\\
 -1-4\ii&1+2\ii&1&3+\ii\\
 2+3\ii&-4-\ii&3-\ii&-2
 \end{pmatrix}.
\end{equation}
Since $M=M^\dagger$, the Cayley transform is exactly unitary. The same $C$ is used at every size.

For sorted energies $E_n$, define
\begin{equation}\label{eq:ratio}
 \delta_n=E_{n+1}-E_n,\qquad
 r_n=\frac{\min(\delta_n,\delta_{n+1})}{\max(\delta_n,\delta_{n+1})}.
\end{equation}
For a slowly varying spectral density, these ratios largely cancel the local density scale, avoiding the need for unfolding.
We use the middle $80\%$ of each energy sequence. More precisely, for $\mathcal N$ sorted levels indexed from zero, we retain $E_{\lfloor\mathcal N/10\rfloor},\ldots,E_{\lceil9\mathcal N/10\rceil-1}$ and form ratios from consecutive gaps within this window.

\subsection{The \texorpdfstring{$N_B=1$}{NB=1} sector}\label{sec:one-statistics}
We take $N_B=1$ and physical momentum $k=0$. As established in Sec.~\ref{sec:real-space-one-label}, this block has dimension $2^L$.

We obtain the energies from Eq.~\eqref{eq:cosine} by diagonalizing the $2^L\times2^L$ unitary color matrix $\cD_L=T_{\rm col}C_{L-1,L}^\dagger$ defined in Eq.~\eqref{eq:explicit-color-transport}. Its eigenvalues $e^{\ii\theta_a}$ determine all $2^L$ energies in this momentum block. We first examine the phases themselves, then ask how their correlations appear in the energies after the cosine map.

For the color gate~\eqref{eq:numerical-gate}, we find numerically that the $8192$ phases at $L=13$ populate the full unit circle. Dividing the circle into $16$ equal arcs gives between $511$ and $514$ phases in each arc, close to the uniform-density value $512$. The largest gap between consecutive phases is $0.002421$ radians. Thus there is no large empty arc, and the density is nearly uniform on scales containing many levels.

To measure the local phase correlations, we order the $\mathcal N=2^L$ phases in $[-\pi,\pi)$ and form all consecutive angular gaps, including the gap that closes the circle. We rescale them by their mean $2\pi/\mathcal N$:
\begin{equation}\label{eq:circular-phase-spacings}
\begin{aligned}
 g_a&=\theta_{a+1}-\theta_a,\qquad a=1,\ldots,\mathcal N,\\
 \theta_{\mathcal N+1}&=\theta_1+2\pi,\qquad
 s_a=\frac{\mathcal N}{2\pi}g_a.
\end{aligned}
\end{equation}
Every gap is counted once, and $\langle s\rangle=1$. The approximately uniform phase density motivates this constant rescaling. We use the complete circle, without the energy-window restriction introduced above.

\begin{figure}[!tbp]
\centering
\includegraphics[width=0.82\textwidth]{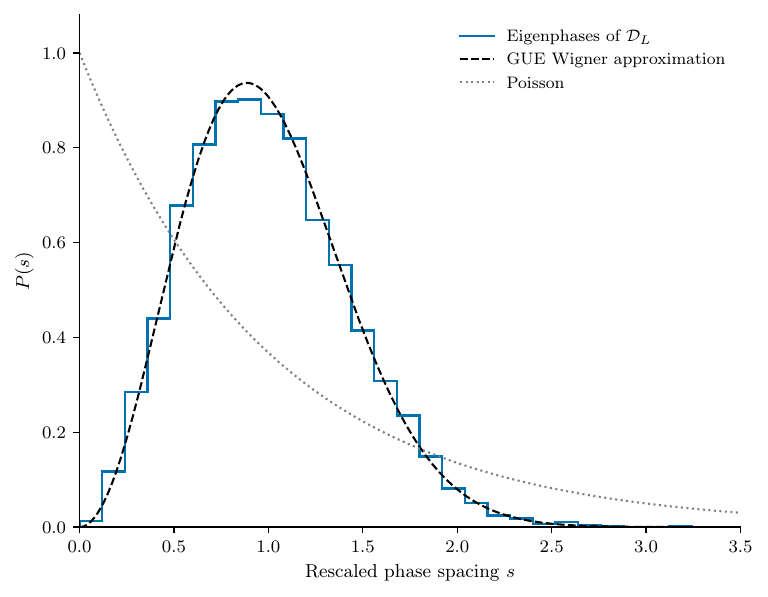}
\caption{Consecutive eigenphase spacings of $\cD_L$ for $L=13$ and the color gate~\eqref{eq:numerical-gate}. All $8192$ circular gaps, including the closing gap, are rescaled to unit mean using~\eqref{eq:circular-phase-spacings}. The distribution shows level repulsion close to the GUE Wigner approximation~\eqref{eq:wigner-unitary-spacing}; the Poisson reference is shown for comparison.}\label{fig:phase-spacings}
\end{figure}

The eigenphases of the unitary operator $\cD_L$ invite comparison with the circular unitary ensemble (CUE). Its local correlations have the same large-matrix limit as those of the Gaussian unitary ensemble (GUE)~\cite{Nishigaki2026CUE}. For the spacing distribution, we use the Wigner approximation to this common limiting law~\cite{AtasBogomolnyGiraudRoux2013}:
\begin{equation}\label{eq:wigner-unitary-spacing}
 P_{\rm W}(s)=\frac{32}{\pi^2}s^2e^{-4s^2/\pi}.
\end{equation}
Figure~\ref{fig:phase-spacings} shows phase repulsion consistent with this unitary reference. At $L=13$, the fraction of spacings with $s<0.2$ is $0.00830$, compared with $0.00839$ from~\eqref{eq:wigner-unitary-spacing} and $0.18127$ from the Poisson reference $P(s)=e^{-s}$. Applying the min/max ratio of Eq.~\eqref{eq:ratio} to consecutive circular gaps, with $g_{\mathcal N+1}=g_1$, gives a mean $0.59434$. This is close to the large-matrix unitary reference $0.5996$, whereas the Poisson value is $2\log2-1=0.38629\ldots$~\cite{AtasBogomolnyGiraudRoux2013}. The mean eigenphase gap ratios $\langle r\rangle_\theta$ for $L=8,\ldots,13$ are presented in Table~\ref{tab:one} below, alongside the energy gap ratios discussed next.

We now return to the energies. At $k=0$, Eq.~\eqref{eq:cosine} maps both $0<\theta<\pi$ and $-\pi<\theta<0$ onto the same energy interval $(L-4,L)$. Two phases far apart on the circle can therefore give neighboring energies. Repulsion between neighboring phases does not by itself imply repulsion between neighboring energies when both branches are combined.

We therefore compare the complete energy sequence with the two sequences obtained by retaining positive and negative principal phases separately.
The conserved charge $Q_3$, obtained from the second logarithmic derivative of the transfer matrix~\eqref{eq:transfer} at $u=0$, also distinguishes the two branches, although we do not use it here.

For the energy-spacing histograms, we use the smooth counting function obtained by mapping a uniform phase density through the cosine dispersion~\eqref{eq:cosine}. For a sequence containing $\mathcal N$ levels before window selection and band midpoint $E_c=L-2$, this gives
\begin{equation}\label{eq:arcsine-unfolding}
\begin{split}
 \overline{\mathcal N}(E)
 &=\mathcal N\left[1-\frac{1}{\pi}\arccos\!\left(\frac{E-E_c}{2}\right)\right],\\
 &\hspace{3em}|E-E_c|\leq2.
\end{split}
\end{equation}
The finite-size phase density need not be exactly uniform; Eq.~\eqref{eq:arcsine-unfolding} specifies our smooth unfolding prescription. It accounts for the accumulation of energies near the band endpoints. Within the central window of each energy sequence, we form the differences $\overline{\mathcal N}(E_{n+1})-\overline{\mathcal N}(E_n)$ and divide by their mean to obtain the unfolded spacings $s$.

\begin{figure}[!tbp]
\centering
\includegraphics[width=0.82\textwidth]{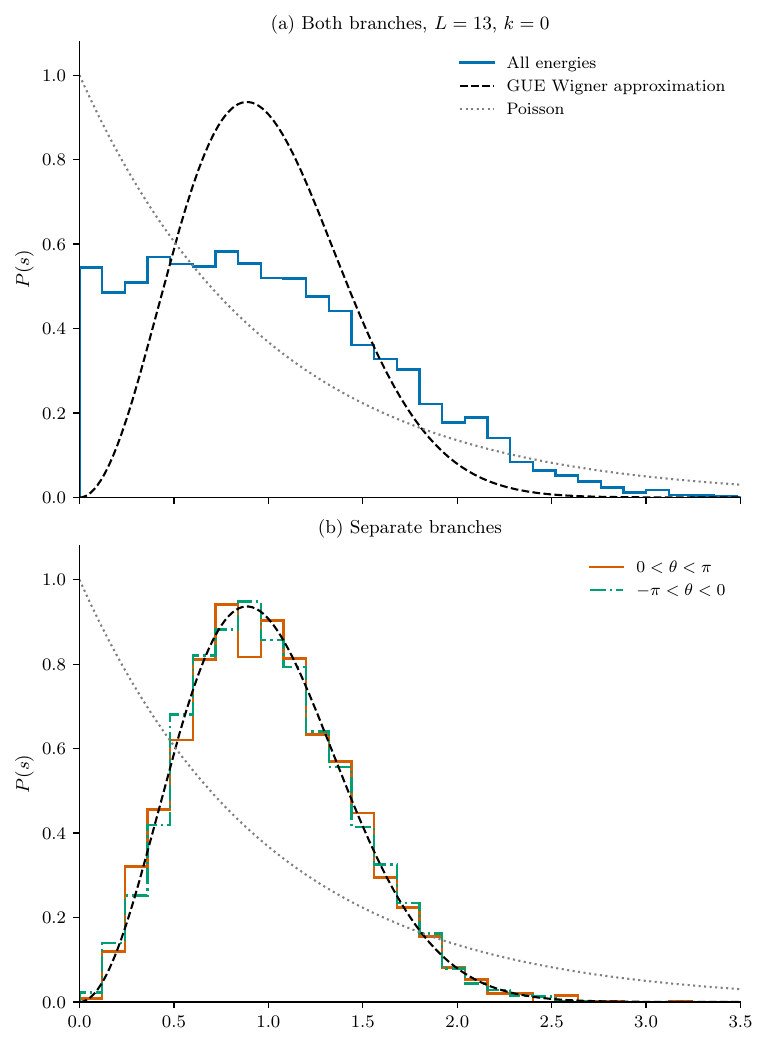}
\caption{Energy-spacing distributions for $N_B=1$, $k=0$, $L=13$, and the color gate~\eqref{eq:numerical-gate}. (a) Both phase branches combined. (b) Energies from positive and negative principal phases separately. Each energy sequence is restricted to its central $80\%$, unfolded with~\eqref{eq:arcsine-unfolding}, and rescaled to unit mean spacing. The separate branches show repulsion close to the GUE Wigner approximation (dashed); the Poisson reference is dotted.}\label{fig:one}
\end{figure}

Figure~\ref{fig:one} confirms this distinction. Combining the branches suppresses level repulsion, while the two separate energy sequences follow the unitary reference much more closely. Table~\ref{tab:one} compares their mean gap ratios with the circular phase statistic. At $L=13$, the full energy sequence gives $0.42034$, close to the reference $0.422$ for a superposition of two equal-density GUE sequences~\cite{GiraudMaceVernierAlet2022}. The positive and negative phase branches give $0.59199$ and $0.58882$.

\begin{table}[!tbp]
\caption{Phase and energy gap ratios for $N_B=1$ and $k=0$. The column $\langle r\rangle_\theta$ uses all circular phase gaps of $\cD_L$. The remaining columns use the central $80\%$ of the energy sequences: all phases together ($0$), positive phases ($+$), and negative phases ($-$). Each full momentum block has dimension $2^L$.}\label{tab:one}
\centering
\begin{tabular}{rrrrr}
\toprule
$L$ & $\langle r\rangle_\theta$ & $\langle r\rangle_0$ & $\langle r\rangle_+$ & $\langle r\rangle_-$\\
\midrule
8 & 0.57469 & 0.46368 & 0.56314 & 0.61215\\
9 & 0.59907 & 0.42983 & 0.60187 & 0.59716\\
10 & 0.60127 & 0.41997 & 0.62210 & 0.57294\\
11 & 0.59155 & 0.41975 & 0.59067 & 0.59341\\
12 & 0.60001 & 0.42562 & 0.59607 & 0.59833\\
13 & 0.59434 & 0.42034 & 0.59199 & 0.58882\\
\bottomrule
\end{tabular}
\end{table}

\subsection{Half filling at physical momentum \texorpdfstring{$k=0$}{k=0}}\label{sec:half}
We now turn to finite label density and examine the half-filled sector $N_A=N_B=L/2$ at physical momentum $k=0$. We obtain this spectrum from the scalar-twist XXX decomposition derived in Sec.~\ref{sec:periodic-general}, Eq.~\eqref{eq:directsum}, using the common color eigenbasis~\eqref{eq:joint-color-eigenbasis}. In each color channel $a$, we retain precisely the label states with twisted translation eigenvalue $t=\eta_a^{-1}$, as required by~\eqref{eq:physical-momentum-twist}. At $L=12$, the physical $k=0$ block contains $315\,516$ energies.

Fixing physical momentum does not remove all degeneracies. For example, at $L=8$ we find $140$ levels at $E=4$. The raw spacing measure at sizes with degeneracies therefore contains a zero-spacing contribution and is \emph{not} a pure Poisson distribution. We report separately the distinct-level sequence obtained by clustering adjacent energies within $\epsilon_E=10^{-10}$, retaining the mean of each cluster.

\begin{table}[tbp]
\caption{Half filling at physical momentum $k=0$, with all color channels retained. $\mathcal N$ is the number of eigenvalues in this momentum block, $\mathcal N_{\epsilon}$ is the number of energy clusters at resolution $10^{-10}$, and $f_0$ is the fraction of adjacent raw gaps not exceeding that resolution. The last column uses the central $80\%$ of the cluster sequence.}\label{tab:half}
\centering
\begin{tabular}{rrrrr}
\toprule
$L$ & $\mathcal N$ & $\mathcal N_{\epsilon}$ & $f_0$ & $\langle r\rangle_{\epsilon}$\\
\midrule
4 & 26 & 21 & 0.2000 & 0.32484\\
6 & 216 & 216 & 0.0000 & 0.34866\\
8 & 2254 & 2115 & 0.0617 & 0.38141\\
10 & 25808 & 24768 & 0.0403 & 0.38767\\
12 & 315516 & 298713 & 0.0533 & 0.38701\\
\bottomrule
\end{tabular}
\end{table}

\begin{figure}[!tbp]
\centering
\includegraphics[width=0.82\textwidth]{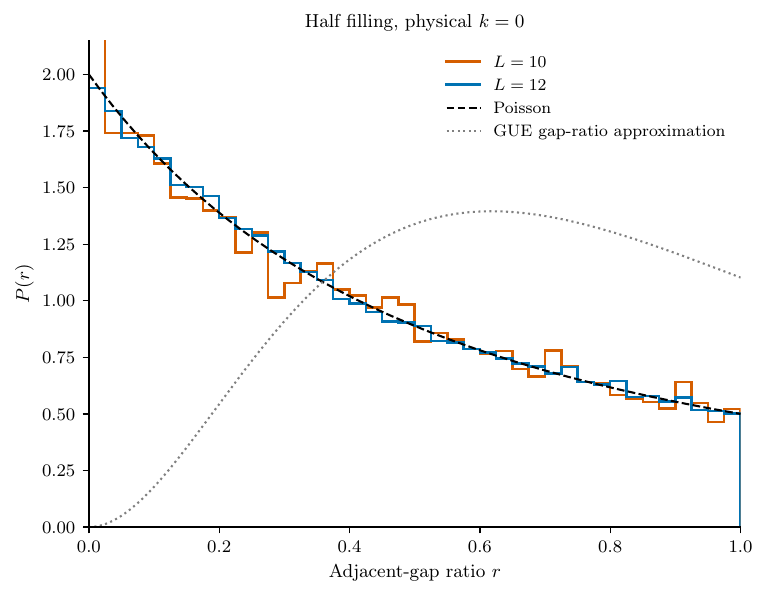}
\caption{Half-filled ratio distributions at physical momentum $k=0$ for the gate~\eqref{eq:numerical-gate}, after energy clustering at $10^{-10}$. All color channels satisfying the momentum condition~\eqref{eq:physical-momentum-twist} are retained. The gap-ratio distributions in the central $80\%$ of the cluster sequences are compared with $P(r)=2/(1+r)^2$, expected for uncorrelated (Poisson) levels, and with the GUE gap-ratio approximation.}\label{fig:half}
\end{figure}

The mean ratios at $L=10$ and $L=12$ are $0.38767$ and $0.38701$, respectively. Figure~\ref{fig:half} compares the full ratio distributions with the Poisson and GUE references. The clustered statistics at these sizes are close to Poisson, although finite-size deviations remain, especially at smaller $L$. For $L=12$, changing $\epsilon_E$ from $10^{-12}$ to $10^{-9}$ changes the central-window mean from $0.38700$ to $0.38709$.

One may ask why the half-filled sector behaves differently from the $N_B=1$ sector, since both require diagonalizing the full color space. The difference lies in the label dynamics that remains after choosing a common color eigenvector and fixing the physical momentum. For $N_B=1$, these choices leave a single label state, whose energy is determined by the color transport phase. At half filling, they instead leave an exponentially large twisted XXX block, with dimension of order $\binom{L}{L/2}/L$. Each color channel therefore contributes an entire many-body spectrum, rather than a single energy. For fixed twist, the label Hamiltonian is the integrable XXX chain, for which Poisson-like spectral correlations are expected. The full momentum-block spectrum superposes exponentially many such sequences without introducing coupling between their states. If correlations between different channels are sufficiently weak, this superposition preserves the absence of level repulsion~\cite{GiraudMaceVernierAlet2022}, providing a physical explanation for the near-Poisson statistics observed here.

\section{Hilbert space fragmentation}\label{sec:fragmentation}\label{sec:commutant}
In this section we establish Hilbert-space fragmentation for both open and periodic boundary conditions.

There are multiple definitions of Hilbert space fragmentation. In~\cite{MoudgalyaMotrunich2022Commutant} fragmentation was defined through the growth of the commutant algebra, consisting of all operators that commute with every local interaction. Conventional symmetries give a commutant dimension that grows at most polynomially with the system size. In the definition adopted here, Hilbert-space fragmentation occurs when this dimension grows exponentially with the system size.

Throughout this section we work in a fixed particle-number sector $\cH_{L,N_B}$ with $0<N_B<L$. We write $\Omega$ for the color holonomy $\Omega_{L,N_B}$ in this sector. Following the definition of fragmentation, we introduce the open- and periodic-boundary commutants on this space as
\begin{equation}\label{eq:bond-commutants}
\begin{aligned}
 \mathcal C_{\rm o}&=\{Q:[Q,R_j]=0,\ 1\leq j<L\},\\
 \mathcal C_{\rm p}&=\{Q:[Q,R_j]=0,\ 1\leq j\leq L\}.
\end{aligned}
\end{equation}

For open boundaries, the sorting transformation $G$ from~\eqref{eq:open-many-gauge} lets us construct conserved operators from arbitrary color matrices.

\begin{theorem}[Open-boundary common commutant]\label{thm:commutant}\label{thm:open-commutant}
Let $C\in U(4)$, and let $G$ be the corresponding sorting transformation. The operators
\begin{equation}\label{eq:dressed-color-operator}
 Q_X=G(I\otimes X)G^\dagger,
 \qquad X\in\operatorname{End}(\mathcal F),
\end{equation}
form a subalgebra of $\mathcal C_{\rm o}$ isomorphic to $\operatorname{End}(\mathcal F)$. Hence
\begin{equation}\label{eq:open-commutant-growth}
 \dim\mathcal C_{\rm o}\geq4^L.
\end{equation}
\end{theorem}
\begin{proof}
Equation~\eqref{eq:bulk} maps every open bond to a label permutation tensored with $I$. Hence $I\otimes X$ commutes with every transformed bond, so $Q_X$ commutes with every $R_j$ for $1\leq j<L$. Since $\dim\mathcal F=2^L$, the embedded algebra has dimension $4^L$, proving the bound.
\end{proof}

For periodic boundaries, the transformed closing bond~\eqref{eq:sortedH} contains the nonlocal color unitary $\Omega$ and its adjoint. A dressed color operator~\eqref{eq:dressed-color-operator} commutes with this bond precisely when its color part commutes with $\Omega$. We therefore define
\begin{equation}\label{eq:color-holonomy-commutant}
 \mathcal A_\Omega
 =\{X\in\operatorname{End}(\mathcal F):[X,\Omega]=0\}.
\end{equation}
The product formula~\eqref{eq:Omega} specifies $\Omega$, but does not supply its eigenbasis for generic $C$. Constructing a basis of $\mathcal A_\Omega$ through the spectral decomposition of $\Omega$ therefore leaves a many-qubit problem. We can nevertheless establish a lower bound on the dimension of $\mathcal A_\Omega$ without constructing an explicit basis of this operator algebra.

\begin{theorem}[Periodic-boundary common commutant]\label{thm:periodic-commutant}
For every $C\in U(4)$ and $0<N_B<L$, the map $X\mapsto Q_X$ embeds $\mathcal A_\Omega$ into $\mathcal C_{\rm p}$. Consequently,
\begin{equation}\label{eq:commutant-growth}
 \dim\mathcal C_{\rm p}\geq2^L.
\end{equation}
\end{theorem}
\begin{proof}
Equation~\eqref{eq:bulk} gives commutation with every bulk bond, and $X\in\mathcal A_\Omega$ gives commutation with the closing bond. Since $\Omega$ is unitary on the $2^L$-dimensional color space, it has an orthonormal eigenbasis. The $2^L$ rank-one projectors onto these eigenvectors are linearly independent elements of $\mathcal A_\Omega$, proving the bound.
\end{proof}

For generic $C$, the conserved operators $Q_X$ are nonlocal. This nonlocality comes from the sorting transformation $G$. For generic $G$, we have not found a description of the commutant algebra in terms of local operators. We therefore refer to this structure as nonlocal Hilbert-space fragmentation.

With open boundaries, the commutant contains the full color matrix algebra. Besides projectors onto invariant color channels, it includes conserved operators mapping one channel to another. These operators relate identical copies of the label dynamics and enforce at least $2^L$-fold degeneracy of every energy level, despite their generally nonlocal physical action.

With periodic boundaries, the closing bond restricts the conserved color operators to those commuting with the holonomy, so different channels need not remain degenerate. The surviving algebra is nevertheless exponentially large: even at fixed particle number, the Hilbert space splits into exponentially many invariant channels, each carrying a twisted XXX Hamiltonian. At $N_B=1$ and fixed physical momentum, each channel contributes only one state, so the observed Wigner--Dyson-like correlations occur between energies belonging to different fragments. These correlations therefore do not by themselves establish chaotic dynamics within the particle-number sector.

\section{Conclusion}\label{sec:conclusions}
We have constructed a homogeneous nearest-neighbor Yang--Baxter-integrable spin-chain family with fixed local dimension and an arbitrary $4\times4$ color unitary. With open boundaries, a nonlocal transformation removes the color action of $C$, mapping the Hamiltonian to an XXX chain tensored with the identity on the color space. With periodic boundaries, an operator-valued twist remains. The resulting spectra can exhibit correlations characteristic of chaotic models, as observed numerically in the $N_B=1$ sector.

The model exhibits Hilbert-space fragmentation with both boundary conditions. The open-chain commutant contains the full color algebra and enforces exponential energy degeneracies. For periodic boundaries, our numerical results at $N_B=1$ and $k=0$ show level repulsion close to the unitary random-matrix prediction when the two branches of the cosine dispersion are analyzed separately. Since each invariant fragment contributes only one state to this momentum block, the observed spectral correlations link different fragments and do not imply chaotic mixing within the particle-number sector.

The relation to the circuit studied in Ref.~\cite{Pozsgay2026SquareRoot} also motivates investigating operator-entanglement growth under the Hamiltonian dynamics considered here. We hope to return to this problem in future work.

\section*{Acknowledgements}
This work was supported by the National Research, Development and Innovation Office (NKFIH, Hungary)
under Grants K-145904 and 2026-4.1.1-MISSZI\'OK-2026-00005.

\appendix
\section{Algebraic properties of the local gate}\label{app:local-algebra}
The integrability construction in Sec.~\ref{sec:YB} requires the local gates to be Hermitian involutions satisfying the braid relation. Physically, a rightward hop of a $B$ label applies $C$ to the colors, while a leftward hop applies $C^\dagger$. We now prove that unitarity of $C$ suffices for these properties. We first work on two sites, then compare the two sequences of three adjacent exchanges for every label configuration. We next derive the Yang--Baxter equation for the Baxterization. In Appendix~\ref{app:periodic-translation}, we will also derive the action of physical translation after the nonlocal change of basis.

\subsection{Hermiticity and involutivity}
In the block-matrix form~\eqref{eq:gate-block-matrix}, taking the adjoint exchanges the two off-diagonal blocks and conjugates each of them, so $R^\dagger=R$. This step uses the adjoint pairing in the definition of the local gate and does not require unitarity.

Squaring the block matrix returns each mixed-label state to its original label sector. The resulting color maps are
\begin{equation}
 R^2=\operatorname{diag}(I_4,CC^\dagger,C^\dagger C,I_4)=I_{16}.
\end{equation}
The last equality uses $C^\dagger C=CC^\dagger=I_4$. Thus two successive exchanges restore both labels and colors. Without unitarity, Hermiticity would remain valid, but involutivity would fail in the mixed-label sectors. Embedding these two-site identities into the chain proves the first two statements in~\eqref{eq:localalgebra}. Local gates on disjoint bonds commute because they act on separate tensor factors.

\subsection{Braid relation}
It remains to compare $R_1R_2R_1$ and $R_2R_1R_2$ on three consecutive sites. We write a state as $\ket{w}\otimes\chi$, where $w$ is a three-letter label word and $\chi\in(\mathbb C^2)^{\otimes3}$ is an arbitrary color vector in physical site order. Here $R_1$ acts on sites $(1,2)$ and $R_2$ on $(2,3)$; products act from right to left.

For example, the first sequence on $BAA$ moves the $B$ label two sites to the right:
\begin{equation}
\begin{split}
 \ket{BAA}\otimes\chi
 &\xrightarrow{R_1}\ket{ABA}\otimes C_{1,2}\chi\\
 &\xrightarrow{R_2}\ket{AAB}\otimes C_{2,3}C_{1,2}\chi\\
 &\xrightarrow{R_1}\ket{AAB}\otimes C_{2,3}C_{1,2}\chi.
\end{split}
\end{equation}
In the second sequence, $R_2$ first acts on $AA$ and hence is the identity; $R_1$ and then $R_2$ perform the same two rightward hops. The order of the color matrices therefore agrees without requiring $C_{1,2}$ and $C_{2,3}$ to commute.

Applying the same rule to all eight input words gives the following output word and color map on each side:
\begin{equation}\label{eq:braid-cases}
\setlength{\arraycolsep}{3pt}
\begin{array}{c c c c}
 w & w_{\rm out} & R_1R_2R_1 & R_2R_1R_2\\ \hline
 AAA&AAA&I&I\\
 AAB&BAA&C_{1,2}^\dagger C_{2,3}^\dagger&C_{1,2}^\dagger C_{2,3}^\dagger\\
 ABA&ABA&C_{1,2} C_{1,2}^\dagger&C_{2,3}^\dagger C_{2,3}\\
 ABB&BBA&C_{2,3}^\dagger C_{1,2}^\dagger&C_{2,3}^\dagger C_{1,2}^\dagger\\
 BAA&AAB&C_{2,3} C_{1,2}&C_{2,3} C_{1,2}\\
 BAB&BAB&C_{1,2}^\dagger C_{1,2}&C_{2,3} C_{2,3}^\dagger\\
 BBA&ABB&C_{1,2} C_{2,3}&C_{1,2} C_{2,3}\\
 BBB&BBB&I&I
\end{array}
\end{equation}
For $ABA$ and $BAB$, unitarity makes both color maps the identity. In every other row, the ordered products already agree. Since the color vector $\chi$ is arbitrary and the eight label sectors span the three-site space, the two operators coincide. Translation of this identity proves the braid relation in~\eqref{eq:localalgebra}.

Only the alternating-label cases require cancellation of an exchange with its inverse. With two label types, every three-site word either has equal neighboring labels or becomes such a word after one exchange. A third label type would introduce three-distinct-label configurations, which are not covered by this proof and could impose additional constraints on the mixed color maps. For the present model, the braid relation therefore constrains neither the eigenvalues nor the entangling action of the unitary $C$.

\subsection{Yang--Baxter equation}\label{app:local-baxterization}
We now verify that the algebraic relations of the local gate give the spectral-parameter identity~\eqref{eq:braided}. Set $X=R_1$ and $Y=R_2$ on the three-site space. Substituting $\check\cR(u)=I+uR$ into the two sides and subtracting yields
\begin{equation}
\begin{split}
 &(I+uX)(I+(u+v)Y)(I+vX)\\
 &\quad -(I+vY)(I+(u+v)X)(I+uY)\\
 &=uv(X^2-Y^2)+uv(u+v)(XYX-YXY)=0.
\end{split}
\end{equation}
Involutivity cancels the quadratic term, and the braid relation cancels the cubic term. This proves the Yang--Baxter equation for arbitrary complex $u,v$. Multiplication by the full site permutation gives the matrix $\cR(u)$ used in the periodic transfer matrix~\eqref{eq:transfer}; no further condition on $C$ enters this step.

\subsection{Physical translation in the transformed basis}\label{app:periodic-translation}
The XXX blocks of Sec.~\ref{sec:periodic-general} determine the energies, but assigning a physical momentum also requires the color motion accompanying an overall translation. Using the same configuration-dependent bases $U_w$, we derive the color maps~\eqref{eq:translation-color-cases}, the color transport identities~\eqref{eq:general-transport-relations}, and the translation formula in each color channel~\eqref{eq:translation-color-channel}. We work at fixed $L,N_B$ and use the notation $\Omega$, $\cD$, and $\widetilde T$ introduced there.

For a configuration $w$, write $w'=T_{\rm lab}w$. We start from the color map $K_w=U_{w'}^\dagger T_{\rm col} U_w$ in~\eqref{eq:configuration-translation-color}. For $1\leq j\leq L-2$, homogeneity gives $TR_j=R_{j+1}T$. Transforming both bonds by~\eqref{eq:bulk} shows that $K_{s_jw}=K_w$, where $s_j$ exchanges the labels at $j,j+1$. These exchanges permute the first $L-1$ labels and connect every configuration with a fixed final label. Thus $K_w$ depends only on whether $w$ ends in $A$ or $B$.

If $w$ ends in $B$, evaluate $K_w$ on $w_0=A^{N_A}B^{N_B}$. Its translate is $BA^{N_A}B^{N_B-1}$, reached by moving the first $B$ left across the $A$ block. Since $U_{w_0}=I$, substitution into~\eqref{eq:configuration-translation-color} gives
\begin{equation}\label{eq:translation-last-B}
 K_B=(C_{N_A,N_A+1}\cdots C_{1,2})T_{\rm col}.
\end{equation}
For a configuration $w$ ending in $A$, the reference calculation in~\eqref{eq:general-color-transport} gives $K_A=\cD$.
Color translation shifts both ordered site indices by one. Using this identity and~\eqref{eq:Omega} gives
\begin{equation}\label{eq:translation-seam-color}
 K_B=\Omega\cD.
\end{equation}
These two cases prove~\eqref{eq:translation-color-cases} and determine the full transformed translation:
\begin{equation}\label{eq:transformed-physical-translation}
 \widetilde T=(T_{\rm lab}\otimes I)\bigl[(I-n^{\rm label}_{B,L})\otimes\cD+n^{\rm label}_{B,L}\otimes\Omega\cD\bigr].
\end{equation}

We next check that the two color operators commute. Substitution of their gate products, cancellation of inverse gates, and translation of the physical site indices give
\begin{equation}\label{eq:transport-twist-commutation-proof}
\begin{aligned}
 \cD^\dagger\Omega\cD
 &=C_{N_A,N_A+1}(C_{N_A+1,N_A+2}\cdots C_{L-1,L})\\
 &\quad\times(C_{N_A-1,N_A}\cdots C_{1,2})C_{L,1}
 =\Omega.
\end{aligned}
\end{equation}
The two middle products act on the disjoint site sets $\{N_A+1,\ldots,L\}$ and $\{1,\ldots,N_A\}$, so they can be interchanged to obtain the last line. Empty products cover $N_A=1$ and $N_A=L-1$. This proves the commutator identity in~\eqref{eq:general-transport-relations}.

Now apply~\eqref{eq:transformed-physical-translation} $L$ times. Every label crosses the closing bond once: each step contributes $\cD$, and each of the $N_B$ labels $B$ contributes one additional $\Omega$. Since these operators commute, the accumulated color factor is $\cD^L\Omega^{N_B}$. The identity $T^L=I$ proves the second relation in~\eqref{eq:general-transport-relations}. Substitution of the common eigenbasis~\eqref{eq:joint-color-eigenbasis} into~\eqref{eq:transformed-physical-translation} then proves~\eqref{eq:translation-color-channel}.


\begin{thebibliography}{10}
\providecommand{\url}[1]{\texttt{#1}}
\providecommand{\urlprefix}{URL }
\expandafter\ifx\csname urlstyle\endcsname\relax
  \providecommand{\doi}[1]{doi:\discretionary{}{}{}#1}\else
  \providecommand{\doi}{doi:\discretionary{}{}{}\begingroup
  \urlstyle{rm}\Url}\fi
\providecommand{\eprint}[2][]{\url{#2}}

\bibitem{DAlessioKafriPolkovnikovRigol2016}
L.~D'Alessio, Y.~Kafri, A.~Polkovnikov and M.~Rigol,
\newblock \emph{From quantum chaos and eigenstate thermalization to statistical
  mechanics and thermodynamics},
\newblock Advances in Physics \textbf{65}(3), 239 (2016),
\newblock \doi{10.1080/00018732.2016.1198134},
\newblock \eprint{1509.06411}.

\bibitem{BorgonoviIzrailevSantosZelevinsky2016}
F.~Borgonovi, F.~M. Izrailev, L.~F. Santos and V.~G. Zelevinsky,
\newblock \emph{Quantum chaos and thermalization in isolated systems of
  interacting particles},
\newblock Physics Reports \textbf{626}, 1 (2016),
\newblock \doi{10.1016/j.physrep.2016.02.005},
\newblock \eprint{1602.01874}.

\bibitem{BertiniColluraDeNardisFagotti2016}
B.~Bertini, M.~Collura, J.~De~Nardis and M.~Fagotti,
\newblock \emph{Transport in out-of-equilibrium {XXZ} chains: Exact profiles of
  charges and currents},
\newblock Physical Review Letters \textbf{117}(20), 207201 (2016),
\newblock \doi{10.1103/PhysRevLett.117.207201},
\newblock \eprint{1605.09790}.

\bibitem{CastroAlvaredoDoyonYoshimura2016}
O.~A. Castro-Alvaredo, B.~Doyon and T.~Yoshimura,
\newblock \emph{Emergent hydrodynamics in integrable quantum systems out of
  equilibrium},
\newblock Physical Review X \textbf{6}(4), 041065 (2016),
\newblock \doi{10.1103/PhysRevX.6.041065},
\newblock \eprint{1605.07331}.

\bibitem{CauxMossel2011}
J.-S. Caux and J.~Mossel,
\newblock \emph{Remarks on the notion of quantum integrability},
\newblock Journal of Statistical Mechanics: Theory and Experiment
  \textbf{2011}, P02023 (2011),
\newblock \doi{10.1088/1742-5468/2011/02/P02023},
\newblock \eprint{1012.3587}.

\bibitem{Faddeev1996ABA}
L.~D. Faddeev,
\newblock \emph{How algebraic {Bethe Ansatz} works for integrable model}
  (1996), \eprint{hep-th/9605187}.

\bibitem{BerryTabor1977}
M.~V. Berry and M.~Tabor,
\newblock \emph{Level clustering in the regular spectrum},
\newblock Proceedings of the Royal Society of London A \textbf{356}, 375
  (1977),
\newblock \doi{10.1098/rspa.1977.0140}.

\bibitem{BohigasGiannoniSchmit1984}
O.~Bohigas, M.~J. Giannoni and C.~Schmit,
\newblock \emph{Characterization of chaotic quantum spectra and universality of
  level fluctuation laws},
\newblock Physical Review Letters \textbf{52}, 1 (1984),
\newblock \doi{10.1103/PhysRevLett.52.1}.

\bibitem{BenetLeyvrazSeligman2003}
L.~Benet, F.~Leyvraz and T.~H. Seligman,
\newblock \emph{{Wigner--Dyson} statistics for a class of integrable models},
\newblock Physical Review E \textbf{68}, 045201(R) (2003),
\newblock \doi{10.1103/PhysRevE.68.045201},
\newblock \eprint{nlin/0306009}.

\bibitem{RelanoDukelskyGomezRetamosa2004}
A.~Rela{\~n}o, J.~Dukelsky, J.~M.~G. G{\'o}mez and J.~Retamosa,
\newblock \emph{Stringent numerical test of the {Poisson} distribution for
  finite quantum integrable {Hamiltonians}},
\newblock Physical Review E \textbf{70}, 026208 (2004),
\newblock \doi{10.1103/PhysRevE.70.026208},
\newblock \eprint{nlin/0402011}.

\bibitem{ScaramazzaShastryYuzbashyan2016}
J.~A. Scaramazza, B.~S. Shastry and E.~A. Yuzbashyan,
\newblock \emph{Integrable matrix theory: Level statistics},
\newblock Physical Review E \textbf{94}, 032106 (2016),
\newblock \doi{10.1103/PhysRevE.94.032106},
\newblock \eprint{1604.01691}.

\bibitem{YuzbashyanShastry2013}
E.~A. Yuzbashyan and B.~S. Shastry,
\newblock \emph{Quantum integrability in systems with finite number of levels},
\newblock Journal of Statistical Physics \textbf{150}, 704 (2013),
\newblock \doi{10.1007/s10955-013-0689-9},
\newblock \eprint{1111.3375}.

\bibitem{Pozsgay2026SquareRoot}
B.~Pozsgay,
\newblock \emph{Square-root growth of operator entanglement in an integrable
  brickwork circuit} (2026), \eprint{2609.39401}.

\bibitem{Kumar2009}
B.~Kumar,
\newblock \emph{Exact solution of the infinite-{$U$} {Hubbard} problem and
  other models in one dimension},
\newblock Physical Review B \textbf{79}, 155121 (2009),
\newblock \doi{10.1103/PhysRevB.79.155121},
\newblock \eprint{0812.2977}.

\bibitem{Kumar2013}
B.~Kumar,
\newblock \emph{Exact spin-orbital separation in a solvable model in one
  dimension},
\newblock Physical Review B \textbf{87}, 195105 (2013),
\newblock \doi{10.1103/PhysRevB.87.195105},
\newblock \eprint{1205.6436}.

\bibitem{MajidMarkl1996}
S.~Majid and M.~Markl,
\newblock \emph{Glueing operation for {$R$}-matrices, quantum groups and
  link-invariants of {Hecke} type},
\newblock Mathematical Proceedings of the Cambridge Philosophical Society
  \textbf{119}(1), 139 (1996),
\newblock \doi{10.1017/S0305004100074041},
\newblock \eprint{hep-th/9308072}.

\bibitem{GatevaIvanovaMajid2008}
T.~Gateva-Ivanova and S.~Majid,
\newblock \emph{Matched pairs approach to set theoretic solutions of the
  {Yang--Baxter} equation},
\newblock Journal of Algebra \textbf{319}(4), 1462 (2008),
\newblock \doi{10.1016/j.jalgebra.2007.10.035},
\newblock \eprint{math/0507394}.

\bibitem{Nichita2013}
F.~F. Nichita,
\newblock \emph{{Yang--Baxter} systems, algebra factorizations and braided
  categories},
\newblock Axioms \textbf{2}(3), 437 (2013),
\newblock \doi{10.3390/axioms2030437}.

\bibitem{GomborPozsgay2022}
T.~Gombor and B.~Pozsgay,
\newblock \emph{Superintegrable cellular automata and dual unitary gates from
  {Yang--Baxter} maps},
\newblock SciPost Physics \textbf{12}, 102 (2022),
\newblock \doi{10.21468/SciPostPhys.12.3.102},
\newblock \eprint{2112.01854}.

\bibitem{Maassarani1998XXC}
Z.~Maassarani,
\newblock \emph{The {XXC} models},
\newblock Physics Letters A \textbf{244}(1--3), 160 (1998),
\newblock \doi{10.1016/S0375-9601(98)00322-3},
\newblock \eprint{solv-int/9712008}.

\bibitem{Maassarani1999Multiplicity}
Z.~Maassarani,
\newblock \emph{Multiplicity {$A_m$} models},
\newblock The European Physical Journal B \textbf{7}, 627 (1999),
\newblock \doi{10.1007/s100510050655},
\newblock Erratum: Eur. Phys. J. B 9, 371 (1999),
\newblock \eprint{solv-int/9805009}.

\bibitem{KatsuraMatsuiPalettaPozsgay2025}
H.~Katsura, C.~Matsui, C.~Paletta and B.~Pozsgay,
\newblock \emph{Weak ergodicity breaking with isolated integrable sectors},
\newblock Physical Review Research \textbf{7}, 023099 (2025),
\newblock \doi{10.1103/PhysRevResearch.7.023099},
\newblock \eprint{2412.13951}.

\bibitem{PadmanabhanKorepin2026}
P.~Padmanabhan and V.~Korepin,
\newblock \emph{A universal algebraic construction of {Hilbert}-space fragmentation}
  (2026), \eprint{2611.19470}.

\bibitem{LiebSchultzMattis1961}
E.~Lieb, T.~Schultz and D.~Mattis,
\newblock \emph{Two soluble models of an antiferromagnetic chain},
\newblock Annals of Physics \textbf{16}(3), 407 (1961),
\newblock \doi{10.1016/0003-4916(61)90115-4}.

\bibitem{Nishigaki2026CUE}
S.~M. Nishigaki,
\newblock \emph{Distributions of consecutive level spacings of circular unitary
  ensemble and their ratio: Finite-size corrections and {Riemann} $\zeta$
  zeros},
\newblock Progress of Theoretical and Experimental Physics \textbf{2026}(2),
  023A02 (2026),
\newblock \doi{10.1093/ptep/ptag006},
\newblock \eprint{2507.10193}.

\bibitem{AtasBogomolnyGiraudRoux2013}
Y.~Y. Atas, E.~Bogomolny, O.~Giraud and G.~Roux,
\newblock \emph{Distribution of the ratio of consecutive level spacings in
  random matrix ensembles},
\newblock Physical Review Letters \textbf{110}, 084101 (2013),
\newblock \doi{10.1103/PhysRevLett.110.084101},
\newblock \eprint{1212.5611}.

\bibitem{GiraudMaceVernierAlet2022}
O.~Giraud, N.~Mac{\'e}, {\'E}.~Vernier and F.~Alet,
\newblock \emph{Probing symmetries of quantum many-body systems through gap
  ratio statistics},
\newblock Physical Review X \textbf{12}, 011006 (2022),
\newblock \doi{10.1103/PhysRevX.12.011006},
\newblock \eprint{2008.11173}.

\bibitem{MoudgalyaMotrunich2022Commutant}
S.~Moudgalya and O.~I. Motrunich,
\newblock \emph{{Hilbert} space fragmentation and commutant algebras},
\newblock Physical Review X \textbf{12}, 011050 (2022),
\newblock \doi{10.1103/PhysRevX.12.011050},
\newblock \eprint{2108.10324}.

\end{thebibliography}

\end{document}